\documentclass[
  ]
  {aipproc}

\layoutstyle{8x11double}

\begin{document}

\def\ms{M$_{\odot}$}

\title{The Role of Single Stars of Low and Intermediate Mass  in Galactic Chemical Evolution}

\classification{}
                
\keywords      {Nucleosynthesis; Galactic chemical evolution; AGB stars; s-process}

\author{Nikos Prantzos}{
  address={Institut d' Astrophysique de Paris, 98bis Bd Arago, 75014 Paris}
}

\begin{abstract}
 A short overview is presented of the role that Low and Intermediate mass stars
 play in Galactic Chemical Evolution; their action affects key elements and 
isotopes, like deuterium, $^3$He, $^7$Li, carbon and nitrogen, and s-process nuclei. 
In all those cases, critical uncertainties still remain and are briefly discussed here.
\end{abstract}

\maketitle

%%%%%%%%%%%%%%%%%%%%%%%%%%%%%%%%%%%%%%%%%%%%
%% MAINMATTER
%%%%%%%%%%%%%%%%%%%%%%%%%%%%%%%%%%%%%%%%%%%%

\section{Introduction}

Nucleosynthesis was established as a major astrophysical discipline in the
mid-50ies, after the founding works of Burbidge et al. (1957, B$^2$FH) and A.G.W. Cameron (1957). These were preceeded by a series of observational landmarak papers in the early 50ies, revealing that stars synthesize new elements, either during their hydrostatic evolution or in supernova explosions. The former case is supported by the discovery of radioactive Tc (with a lifetime $\tau\sim$1 Myr for its longest lived isotope $^{99}$Tc) in red giants by Merill (1952), while the latter by the exponential decay with time of the luminosity of supernovae, a clear signature of radioactivity in action (Baade et al. 1954)\footnote{Baade et al. (1956) thought that $^{254}$Cf was at the origin of the supernova luminosity, but in the mid-60ies it was realised that $^{56}$Co was responsible, as finally confirmed by the direct detection of its characteristic gamma-ray lines in SN1987A; both nuclei have similar lifetimes, but $^{56}$Co (the progenitor of stable $^{56}$Fe) is much more abundantly produced in supernova explosions}. 

Another major contribution came through the discovery of two distinct stellar populations in the Milky Way by W. Baade, differing in ages, kinematics and chemical composition: Population I stars, like our Sun, are on average young, metal rich and orbit in the plane of the Galactic disk, while Population II stars are old, metal poor and populate the Galactic halo, having a substantial velocity component perpendicularly to the disk. Those differences in age and metallicity suggest clearly a progressive enrichement  of the galactic gas in metals, through the nucleosynthetic action of successive stellar generations.
Those simple ideas evolved into another asrophysical discipline, closely related to nucleosynthesis, namely Galactic Chemical Evolution (GCE). In GCE models, stars contribute to the enrichment of the interstellar medium (ISM) with heavy elements (metals), but also to the depletion of some light and fragile elements
which are burned in their interiors ({\it astration}). Their contribution depends on their mass $M$, which determines:

i)  the time when the star releases its products in the ISM. Massive stars ($M>$10 \ms) last for less than 20 Myr and, for most GCE applications, they die immediately after their birth ("instantaneous recycling aproximation"). Those of intermediate mass (2-8 \ms) span a larger range of lifetimes (50 to 1000 Myr), while those lighter than 1.5 \ms \ (defined, somewhat arbitrarily, as low mass stars) live more than 2 Gyr. Obviously, the latter category had no time to affect the chemical evolution of the Galactic halo, which lasted for 1-2 Gyr.

ii) the type and amount of the various nuclei synthesized by the star. Low and intermediate mass (LIM) stars  evolve only up to shell He-burning, leaving behind a carbon-oxygen white dwarf remnant.They enrich the ISM with the products of their central and shell H burning (ejected in the red giant stage, after the 1st and 2nd dredge-up) and of shell He burning (ejected in the asymptotic giant branch or AGB phase, after the 3d dredge-up). The former include $^{3,4}$He, $^{13}$C, $^{14}$N, $^{17}$O, while the latter concern $^{12}$C, heavy s-elements and (perhaps) some $^{22}$Ne  and $^{25,26}$Mg.  Note than in the AGB phase, He burning products, like $^{12}$C, mix with protons and may undergo further H-burning if the bottom of the convective envelope reaches sufficiently high temperatures (the so-called Hot Bottom Burning), leading to the production of $^{13}$C and $^{14}$N (Sect. 3) but also (perhaps) $^7$Li (Sect. 2).

\begin{figure}
  \includegraphics[height=.45\textheight]{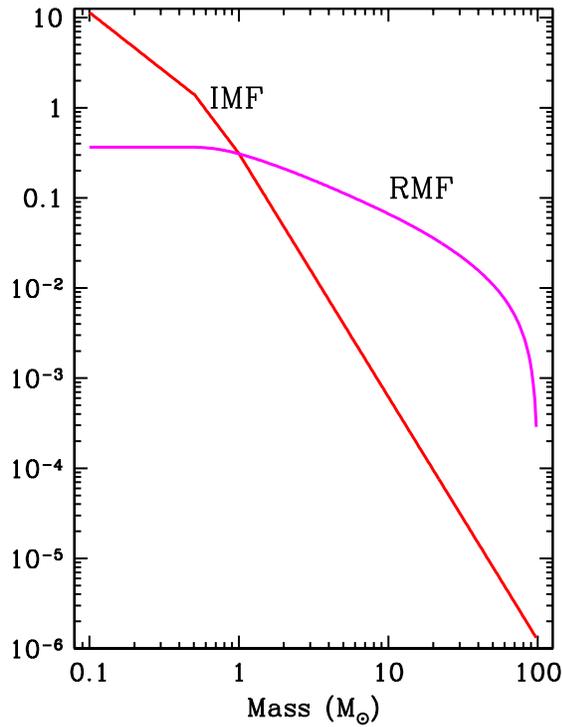}
  \caption{Initial Mass Function (IMF) and Return Mass Fraction (RMF) as a function of stellar mass. The IMF is the one of Kroupa et al. (1993). The RMF(M)
  is given by $R(M) \ = \ \int_M^{100} [M'-M_{REM}(M')] \Phi(M') dM'$. Stars in the mass ranges 1-2, 2-4, 4-9 and 10-100 \ms contribute about equally to the RMF.
  }
\end{figure}

iii) the number of the stars in a particular mass range, which is a decreasing function of mass, the so-called Initial Mass Function (IMF) $\Phi(M)$
$$
\Phi(M) \ = \ {{dN}\over{dM}} \ = \ A \ M^{-(k+1)}
$$

\noindent
with a coefficient $k$ increasing  with mass (i.e. the IMF steepens with mass). The classical value of $k$ (derived by E. Salpeter in 1955 for the 1-10 \ms \ range) is $k$=1.35 and still provides a good fit to modern data, at least up to 50 \ms; lower $k$ values are required at M$<$1 \ms (see e.g. Kroupa 2002 or Chabrier 2003 for recent reviews). For "reasonable" IMFs, the return mass fraction
(RMF)
$$
R \ = \ \int_1^{100} [M-M_{REM}(M)] \Phi(M) dM
$$
is $R\sim$0.33. This is the fraction of the mass of a stellar generation (normalised to unity, i.e. $\int_{0.1}^{100} \Phi(M) MdM $=1) that is returned to the ISM in less than 10 Gyr, i.e. from stars of 1-100 \ms; the rest is blocked mostly  ($\sim$90\%)in lowest mass stars, but also in stellar remnants (white dwarfs, neutrons stars and black holes) of mass $M_{REM}(M)$. The return fraction $R$ is almost equally divided in four parts, corresponding to the mass ranges 1-2, 2-4, 4-9 and 10-100 \ms, respectively, i.e. LIM stars provide about three quarters of the returned mass. This makes them important contributors to the enrichment of the ISM with some heavy elements, and obviously the dominant agents of the astration of fragile elements.

\begin{figure}
  \includegraphics[height=.35\textheight,angle=-90]{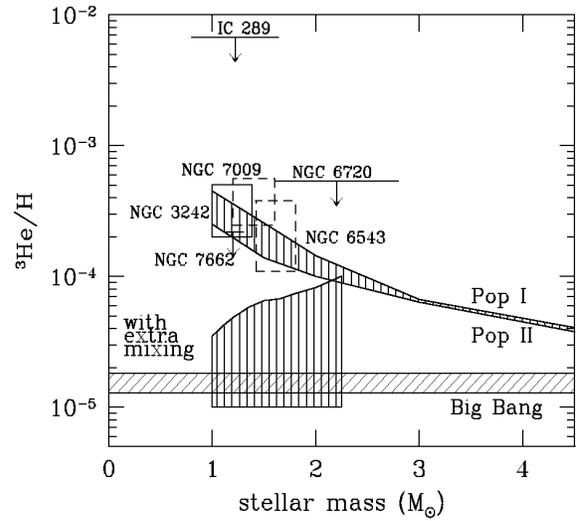}
  \caption{Abundance of $^3$He in Planetary Nebulae (from Galli 2005). {\it Upper shaded aerea} indicates predictions of standard models, in agreement with observations ({\it whithin rectangles}); such high abundances lead to overproduction of $^3$He during galactic evolution (upper curves in Fig. 3). {\it Lower shaded aerea} indicates required level of production in order to avoid overproduction of $^3$He during galactic evolution (lower curves in Fig. 3); such a reduced yield may result from extra-mixing in red giants, also required on other observational grounds (Charbonnel 1995). It should affect 95\% of all stars below 2 \ms, while current observations of $^3$He would concern then the remaining 5\%. 
  }
\end{figure}

In the following we present a short overview of the role played by LIM stars in the chemical evolution of the light isotopes D, $^3$He and $^7$Li, the intermediate mass isotopes $^{12}$C and $^{14}$N, and  the heavy s-process nuclei. 
 
\section{The light isotopes D, $^3$He and $^7$Li}

\begin{figure*} 
\includegraphics[angle=-90,width=0.9\textwidth]{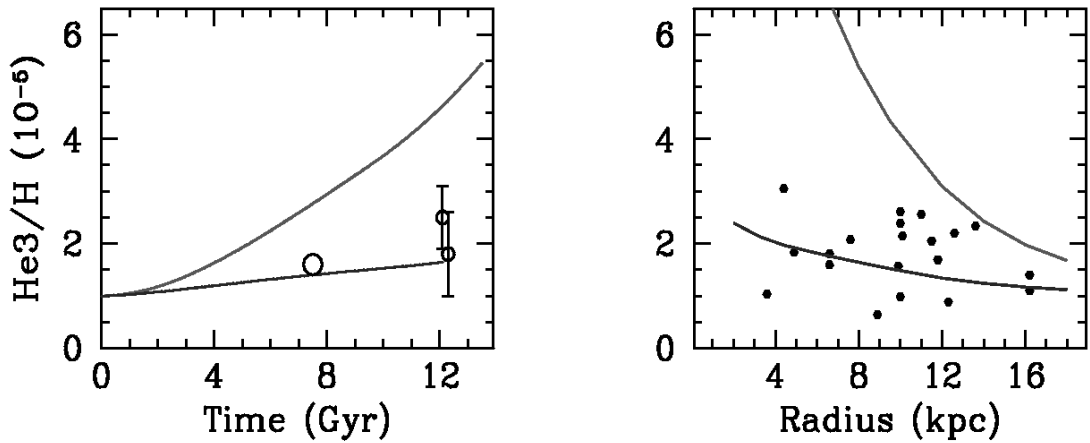}
  \caption{Evolution of the abundance of $^3$He in the solar neighborhood
  as a function of time ({\it left}) and present day profile of $^3$He/H in 
  the Milky Way disk ({\it right}). In both cases, the {\it upper} curves are
  calculated with standard $^3$He yields from LIM stars (and  clearly
  overproduce $^3$He) while the lower ones by assuming that 95\% of the $^3$He of LIM stars is destroyed by some non-standard mechanism; this latter assumption
  allows to satisfy observational constraints, but is not supported by the rare observations of $^3$He in planetary nebulae of presumably known mass (see Fig. 2). In the {\it left panel} pre-solar $^3$He ({\it large circle}) and present day values in the ISM ({\it small circles}) are from Galli (2005). In the {\it right panel}, ISM values are from Bania et al. (2002).
  }  
\end{figure*}

Among the light isotopes produced in standard Big Bang Nucleosynthesis (BBN),
namely D, $^{3,4}$He and $^7$Li, 
\begin{itemize}
\item {one (D) is subsequently depleted by astration,
according to both theory and observations}, and mostly by LIM stars in view of the RMF properties (Sec. 1);
\item{two ($^4$He and $^7$Li) are produced in stars (although it is not yet clear by which stars, LIM or Massive ones, or even novae in the case of $^7$Li), and}
\item{ one ($^3$He) seems not to be affected by GCE (its current abundance being similar to the one resulting from BBN) although
standard stellar models (corroborated by observations of planetary nebulae) suggest that it should.}
\end{itemize}

\subsection{Deuterium}

Modelling the chemical evolution of deuterium is a most straightforward enterprise, since this fragile isotope is 100\% destroyed in stars of all masses
(burning at temperatures $\sim$5 10$^5$ K, already on the pre-main sequence) and has no known source of substantial production other than BBN. If the boundary conditions of its evolution (namely the primordial abundance resulting from BBN and the present day one) were precisely known, the degree of astration, which depends on the adopted IMF and star formation rate, should be severely constrained.

The difficulty to determine the primordial D abundance in the 90ies pushed researchers to turn the problem upside down and try to determine that abundance through reasonable models of local GCE (assuming that the present day abundance is precisely known). Those efforts concluded that reasonable GCE models, reproducing succesfuly the major observational constraints in the solar neihborhood, result only in moderate D depletion, by less than a factor of two (Prantzos 1996, Dearborn et al. 1996, Tosi et al. 1998, Chiappini et al. 2002).

The primordial abundance of D is now well determined (D$_P$/H=2 10$^{-5}$), since observations of D in high redshift gas clouds agree with abundances derived from observations of the Cosmic Microwave Background, combined to standard BBN calculations (e.g. Serpico et al. 2004); it points to negligible D depletion up to solar system formation 4.5 Gyr ago. However, the present day abundance of D in the local ISM is now under debate. Indeed, UV measurements of the FUSE satellite  along various lines of sight suggest substantial differences (a factor of two to three) in D abundance between the Local Bubble (D/H$\sim$1.4 10$^{-5}$ for  $<$100 pc) and beyond it (D/H$\sim$0.5-1 10$^{-5}$ at 100-1000 pc) (see Hebrard et al. 2005). Until the origin of that discrepancy is found, the local GCE of D in the past few Gyr will remain poorly understood (see Geiss et al. 2002).\textit{}

\subsection{$^3$He}

A nice overview of the $^3$He status is recently presented in Galli (2005).
The pre-solar value of $^3$He measured in meteorites ($^3$He/H=1.5 10$^{-5}$, Lodders 2003) is not very different from the primordial one ($^3$He$_P$/H=1 10$^{-5}$, e.g. Serpico et al. 2004 ), inferred from standard BBN and WMAP data analysis. However, since the pioneering work of Iben (1967), stars are known to produce substantial amounts of $^3$He, through the action of p-p chains on the main sequence. The net $^3$He yield varies steeply with mass (roughly $\propto M^{-2}$), since the p-p chains are less effective in more massive stars. In standard stellar models, 1-2 \ms \ stars are the most prolific producers.

Combining those yields with simple GCE models, Rood et al. (1976) found that the pre-solar and present-day abundances of $^3$He are then largely overproduced. The latter are measured either locally by satellite experiments (Gloekler and Geiss 1998) or across the Milky Way disk, through radio observations (Bania et al. 2002). All those measurements point to a current ISM abundance of $^3$He/H$\sim$1.-2 10$^{-5}$, i.e. not very different from the pre-solar value. In other terms, observations show that $^3$He abundance remained $\sim$constant through the ages, while standard stellar models (including recent ones, e.g. Boothroyd and Sackmann 1999) and GCE models (e.g. Prantzos 1996, Romano et al. 2003) point to a large increase (Fig. 3, upper curves).

A possible solution to the problem was suggested by Hogan (1995) and Charbonnel (1995). It postulates destruction of $^3$He in the red giant phase of Low mass stars through some "extra-mixing" mechanism, which brings $^3$He in H-burning zones. The "bonus" is a concommitant modification of the $^{12}$C/$^{13}$C isotopic ratio in red giants, in excellent agreement with observations (Charbonnel and do Nascimento 1998).

Thus, LIM stars should destroy in the red giant phase whatever $^3$He they produce on the main sequence. A possible drawback to the idea is that observations in (at least one) planetary nebulae of known  mass  suggest full agreement with standard model predictions, i.e. with no extra-mixing (see Fig. 2 and Galli 2005). GCE requires that in $>$90\% of the stars, $^3$He produced on the main sequence must be destroyed in the red giant phase, in order to avoid oveproduction (Fig. 3, lower curves). It may well be that current detections of $^3$He in planetary nebulae  concern only the remaining $<$10\% of the stars, but it is still early to draw definitive conclusions.

\subsection{$^7$Li}

Among the two stable Li isotopes, $^7$Li is the most abundant ($^7$Li/$^6$Li=12
in the Sun) and the only one produced in standard BBN. 
The study of its origin and evolution is probably the most complex topic in modern nucleosynthesis. The reason is twofold:

i) the fragility of that isotope, which makes possible that observed abundances in main sequence stars (even "warm" ones, with reduced convective envelopes) are not the true initial ones, and

ii) the fact that, despite that fragility, a multitude of production sites is theoretically possible (a situation not shared by any other isotope).

\begin{figure}
  \includegraphics[height=.45\textheight]{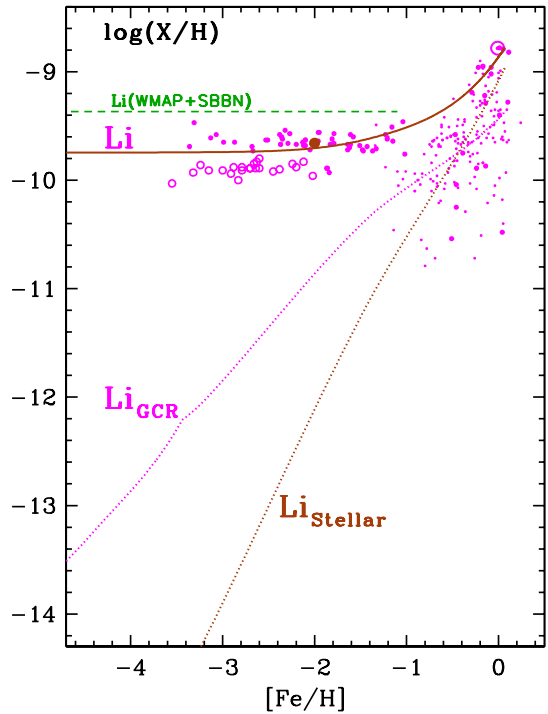}
  \caption{Evolution of the total Li abundance (mostly $^7$Li) with metallicity [Fe/H].  The {\it upper curve (solid) } corresponds to the upper envelope of  "observations" Li$_{\bf Total}$. For halo (low metallicity) stars Li$_{\bf Total}$ here indicates an average between the data of Ryan et al. (1999, {\it open circles}) and those of  Melendez and Ramirez (2004, {\it filled circles}); for high metalicity (disk) stars observations are from Chen et al. (2002). The two {\it dotted curves} indicate the contribution of Galactic Cosmic Rays (Li$_{\bf GCR}$) and stars (Li$_{\bf Stellar}$), respectively. The former represents the sum of $^6$Li and $^7$Li and is derived by requiring that GCR reproduce the observed evolution of $^9$Be (not shown here). The latter is derived as   Li$_{\bf Stellar}$ = Li$_{\bf Total}$-Li$_{\bf GCR}$ and represents stellar production of $^7$Li alone (since stars do not produce $^6$Li). This, empirically derived, stellar component of Li dominates at high metallicities, but it is not clear whether such a strong metallicity dependence can be justified in the framework of realistic stellar models. 
  }
\end{figure}

The cosmological origin of $^7$Li makes no doubt, after the epochal discovery of the Li "plateau" (Spite and Spite 1982) in low metallicity old halo stars.
However, the value of that plateau has been  difficult to establish, because 
of uncertainties in model atmospheres and, in particular, effective temperatures. A factor of two discrepancy exists today (Li/H = 1 - 2 10$^{-10}$,
Ryan et al. 1999, Melendez and Ramirez 2004). But even the highest observed Li plateau values are lower than the one derived from WMAP data on CMB anisotropies combined with standard BBN (Ref. ). This discrepancy may suggest either i) underestimated Li destruction cross-sections in BBN calculations, ii) underestimated systematic errors in Li observations, or iii) systematic Li depletion by at least 0.3 dex in the envelopes of those stars (see Lambert 2004 for an overview of the Li problems).

Whatever the solution of that discrepancy turns out to be, the Li abundance increases steeply with metallicity in disk stars, the solar value being a factor of 4-10 above the primordial one (4 if the WMAP value is assumed as primordial, and 10 if the Ryan et al. plateau value is adopted), suggesting that another Li source has been in operation. Galactic Cosmic Rays (GCR) do synthesize $^7$Li and $^6$Li, along with monoisotopic $^9$Be. The observed evolution of Be (which has no other source than GCR) in the Galaxy allows then to safely predict the corresponding evolution of Li, as shown in Fig. 4. CGR can make only 25\% of the solar Li; the remaining amount has to originate in a stellar source. The evolution of that stellar component is derived in Fig. 4 as Li$_{\bf Stellar}$ = Li$_{\bf Total}$ - Li$_{\bf GCR}$, where Li$_{\bf Total}$ is the upper envelope of observations.   

Several potential stellar sources of $^7$Li have been proposed in the literature
(see Romano et al. 2001):

i) Core collapse supernovae, through neutrino induced nucleosynthesis in the He-shell; neutrinos "spallate" $^4$He and produce mass A=3 nuclei, which interact with $^4$He to produce mass A=7 nuclei. This process is, in principle, independent of metallicity and the resulting Li contribution should appear early on and  scale with metallicity; this does not correspond to the empirically derived Li$_{\bf Stellar}$in Fig. 4.

ii) AGB stars, through Hot Bottom Burning (HBB). In massive AGB stars (4-6 \ms) H-burning may take place if the bottom of the convective envelope is hot enough (T$\sim$50 10$^6$ K); $^3$He+$^4$He produce then unstable $^7$Be which is dragged to the surface and expelled through the stellar wind before decaying into $^7$Li (Cameron-Fowler mechanism). Again, the efficiency of that mechanism should be strongly metallicity dependent in order to reproduce the empirical Li$_{\bf Stellar}$ curve of Fig. 4. Calculations of Ventura et al. (1998) suggest indeed such a dependence, but the absolute Li yields are rather low to match observations (Romano et al. 2001).

iii) Low mass red giants. A small percentage of Pop I red giants display substantial Li abundances, even in excess of the ISM value. The mechanism of that production is still uncertain\footnote{It may be produced by Cool Bottom Process (CBP), as suggested by Boothroyd and Sackmann (1999): $^7$Be is produced in the H-burning shell and rapidly transported (by some ad hoc "extra-mixing" mechanism) to the convective envelope and then to the surface.} as are the duration of the production episode and the yields of Li. In fact, most of that Li should be certainly destroyed shortly after, and before the tip of the red giant branch, by the same process that destroys $^3$He and allows to reproduce observed $^{12}$C/$^{13}$C ratios (see previous section and Charbonnel, these proceedings).
Thus, that mechanism appears of litle importance  for galactic Li production. 

iv) Novae. Those objects, especially those resulting from low mass systems, may certainly meet the constraints of the empirically derived Li$_{\bf Stellar}$ curve (Fig. 4). However, their Li yields (Jos\'e and Hernanz 1998), as well as their frequency of appearance in the Galaxy are very poorly known at present to allow for   a reasonable estimate of their contribution.

In summary, the largest part of solar Li originates from a stellar source which is unknown at present and, most probably, involves LIM stars.

\begin{figure}
  \includegraphics[height=.45\textheight]{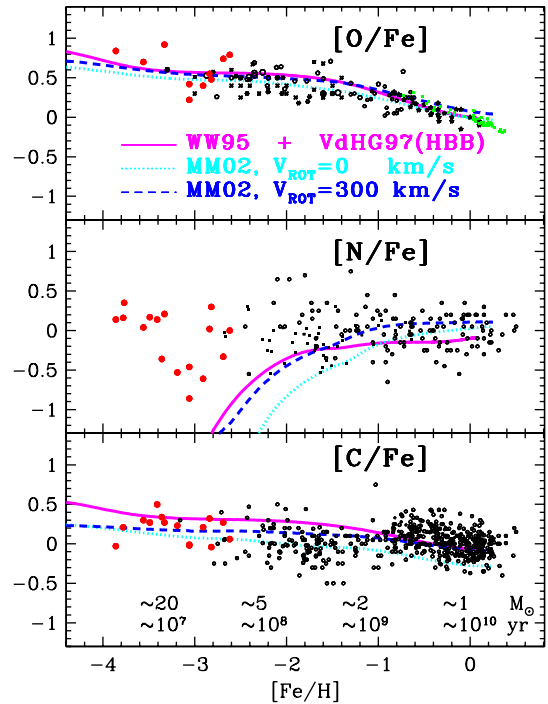}
  \caption{
Evolution of O/Fe ({\it top}), N/Fe ({\it middle}) and C/Fe({\it bottom}) as a  function of [Fe/H]. Curves represent model results with stellar yields from Woosley and Weaver (1995) for massive stars and van den Hoek and Groenewegen (1997) for LIM stars with Hot Bottom Burning ({\it solid}), from  Meynet  and Maeder(2002) without rotation and no HBB for all masses ({\it dashed}) and from Meynet and Maeder (2002) with rotation but no HBB for all masses  ({\it dotted}). In all panels, data a lowest metallicities ({\it thick gray dots}) are from VLT observations of Spite et al. (2004). Rotation helps as much as HBB in producing primary Nitrogen. Numbers in the bottom of the figure indicate approximate timescales of evolution and masses of stars evolving in such timescales. 
  }
\end{figure}

\section {Carbon and Nitrogen}

\begin{figure} 
\includegraphics[height=.45\textheight]{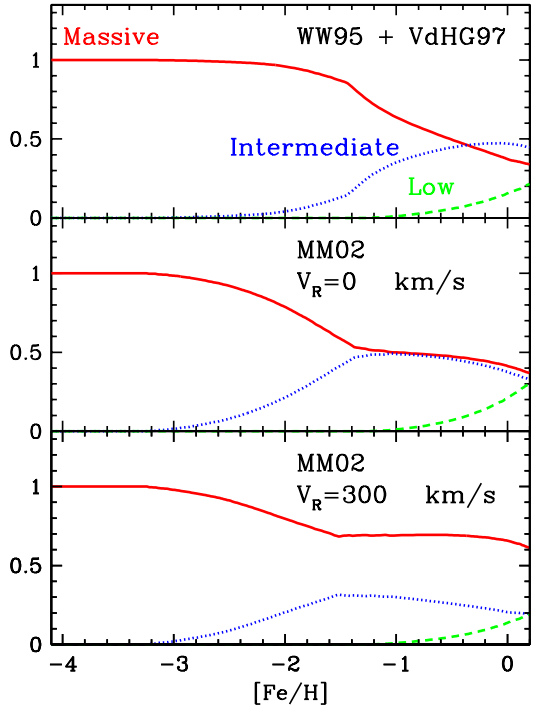}
\caption{Contribution of the various stellar mass ranges to the local galactic production of carbon as a function of [Fe/H]. The three panels display results obtained with yields from from Woosley and Weaver (1995) for massive stars and van den Hoek and Groenewegen (1997) for LIM stars with Hot Bottom Burning ({\it top}), from Meynet and Maeder (2002) without rotation and no HBB for all masses ({\it middle}) and from Meynet and Maeder (2002) with rotation but no HBB for all masses  ({\it bottom}). In all panels, the contributions of Massive stars ($>$10 \ms), Intermediate mass stars (2-9 \ms) and Low mass stars (<1.5 \ms) are indicated by {\it solid}, {\it dotted} and {\it dashed} curves, respectively.
}
\end{figure}
The evolution of CNO elemental abundances with metallicity appears in Fig. 5. C and N faithfully follow Fe (i.e. the C/Fe and N/Fe ratios are always $\sim$solar) while O/Fe is $\sim$3 times solar in low metallicity halo stars and slowly decreases to solar in disk stars. The latter evolution is interpreted as evidence of another, long-lived,  source of Fe in the disk; this  is usually assumed to be type Ia supernova, producing $\sim$2/3 of solar Fe (see Matteucci, these proceedings). This implies that also $\sim$2/3 of solar N and C originate from a long-lived source, which matches quasi-perfectly the late production of Fe from SNIa. 

\begin{figure}
\includegraphics[height=.45\textheight]{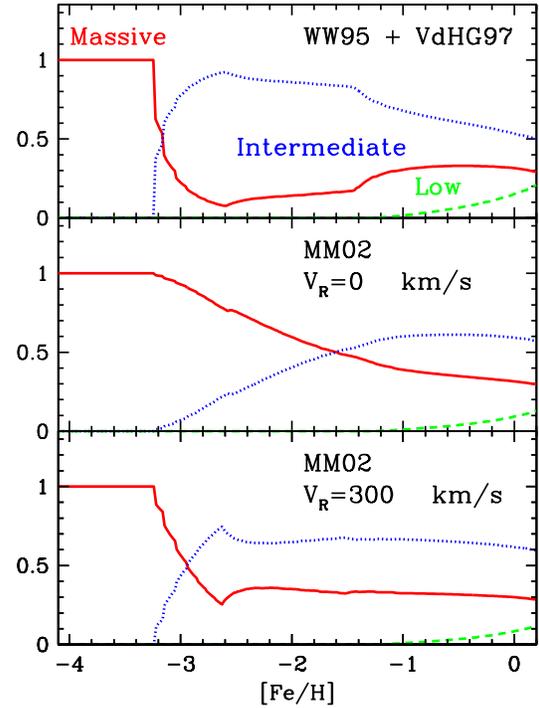}
\caption{Contribution of the various stellar mass ranges to the local galactic production of nitrogen as a function of [Fe/H]. The three panels display results obtained with yields from from Woosley and Weaver (1995) for massive stars and van den Hoek and Groenewegen (1997) for LIM stars with Hot Bottom Burning ({\it top}), from Meynet and Maeder (2002) without rotation and no HBB for all masses ({\it middle}) and from Meynet and Maeder (2002) with rotation but no HBB for all masses  ({\it bottom}). In all panels, the contributions of Massive stars ($>$10 \ms), Intermediate mass stars (2-9 \ms) and Low mass stars (<1.5 \ms) are indicated by {\it solid}, {\it dotted} and {\it dashed} curves, respectively.
}
\end{figure}

That source {\it may} or {\it may not} be LIM stars. Such stars certainly produce C in their He-burning shells, which is dragged to the surface and expelled during the AGB and planetary nebula phase. They also produce N, either in the H-burning shell (as {\it secondary}, from the initial C and N) or, in the case of 3-6 \ms \ AGBs, in the bottom of the convective envelope, through Hot Bottom Burning (this time as {\it primary}, from the C produced in the He-burning shell and dragged in the envelope). However, the absolute yields of C and N as a function of the stellar mass and metallicity are very poorly known in the case of AGBs, because of the various uncertainties affecting the modelisation of that phase (mainly mass loss rates and mixing mechanisms, see e.g. Lattanzio 2004 and Charbonnel, these proceedings).  

In fact, C yields from massive stars, despite their " instantaneous recycling", may   mimic  the behaviour of a delayed source. The reason is the strong dependence of those yields on mass loss, which is a function of metallicity and rotational velocity (e.g. Maeder and Meynet 2002):  massive stars display higher mass losses and larger C yields at high metallicities (although C is always produced as primary). Such metallicity dependent C yields, incorporated in models of GCE, lead to a better (albeit not perfect match, see Prantzos 2003a and Fig. 5) of the late C/Fe behaviour, as originally suggested in Prantzos et al. (1994, see also Gustaffson et al. 1999). Indeed, the fractional contribution of rotating massive stars with mass loss to C production may be larger than the one of LIM stars (bottom panel of Fig. 6).

The case of N is different. Although mass loss favours its release befor subsequent destruction, it is  produced mostly as secondary in massive star H-burning; its yields (even from rotating stars) are never large enough as to dominate its galactic evolution.
Most of it ($\sim$2/3, see Fig. 7) originates from massive AGB stars of 3-6 \ms, at least when stellar yields currently available are adopted (which {\it assume} Hot Bottom Burning to produce N in large quantities). However, N/Fe remains $\sim$solar even at the lowest metallicities (Fig. 5), at such early  times (presumably $<$10$^8$ yr) that  AGB stars had not yet appeared. Either the timescales of early galactic evolution are underestimated (Prantzos 2003b) or massive stars may, after all, produce a lot of primary N through some extra-mixing process of protons in He-burning zones.

In summary, in view of all current uncertainties on stellar yields, it is not yet clear whether the dominant source of C at high metallicities is massive or LIM stars (see Chiappini et al. these proceedings for a different view, clearly in favour of LIM stars). As for N, most of it at high metallicities aparently originates from AGB stars. However, its origin (as a primary element) at very low metallicities, remains a mystery.

\section{Heavy s-elements}

 Most of the heavier than Fe nuclei are produced by neutron captures, and about half of them by the s(slow)-process, which takes place on time scales long w.r.t. typical beta-decay timescales. As shown by Clayton et al. (1961) in the framework of parametrised models, the outcome of that nucleosynthesis depends on the time-integrated neutron exposure, with shorter exposures favoring the production of lighter s-nuclei. The solar system s-abundance distribution
(with the heaviest nuclei being less abundant than lighter ones) suggests that more material has been exposed to short exposures than to long ones.

\begin{figure}
\includegraphics[height=.28\textheight]{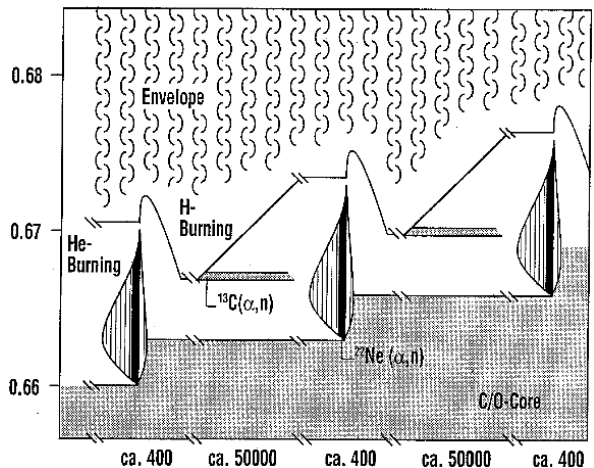}
\caption{Schematic view of the interior of a thermally pulsing AGB star, near the bottom of its convective envelope. The lower shaded aerea  indicates the C-O core (ashes of He-burning). Convective regions (vertically shaded) are produced when He ignites in the border of that aerea, and they mix He-burning products,
like $^{12}$C, up to the inert He region. After the He-flash is over, protons from the H-zone mix (by some still unspecified mechanism) into the now radiative He zone and are captured by $^{12}$C to produce $^{13}$C. The $^{13}$C($\alpha,n$) reaction releases neutrons in a tiny radiative region, which are subsequently captured by Fe-peak nuclei (s-process). The operation of $^{22}$Ne($\alpha,n$) at higher temperatures, in the subsequent He-flash, is also required in order to  fully reproduce the solar system s-nuclei distribution. It should be noted that this scheme is never found in  self-consistent models of AGB stars, but is rather a theoretical consruction (forced upon the model stars) in order to achieve agreement with observations.
}
\end{figure}

Today, it is commonly admitted that s-nuclei up to mass number A=90 are produced in a different site than heavier ones. The corresponding (relatively) short  neutron exposure is obtained in the He-burning cores of massive stars, where neutrons are released through $^{22}$Ne($\alpha,n)$; realistic stellar models succesfully reproduce the solar s-abundance distribution up to A=90 in such a simple environment (e.g. Prantzos et al. 1990). Heavier s-nuclei are produced in lower mass stars, as the observations of Tc by Merill (1952) clearly suggest. However, the corresponding astrophysical environment is less well determined
(see Goriely and Siess 2005 for a recent critical overview of the subject).

 The current paradigm of the synthesis of A$>$90 s-nuclei (the so-called {\it main s-component}) involves Low mass AGB stars ($\sim$1.5 \ms, e.g. Gallino et al. 1998) and a typical setting is schematically presented in Fig. 8. The key underlying idea is the existence of a $^{13}$C "pocket" in a He-rich region, such as the $^{13}$C($\alpha,n$) neutron source operates at not too high temperatures and in a radiative zone. Since $^{13}$C is "regenerated" after each burning episode, a large overall neutron exposure can be achieved (which is not the case with $^{22}$Ne in massive stars).
Paramerized models of that senario have been shown to reproduce satisfactorily
a large body of observational data. Such models assume either i) the presence a suitable amount of $^{13}$C in the appropriate zone (e.g. Gallino et al. 1998, Busso et al. 2001) or ii) instantaneous injection of protons from the envelope in the C-rich region (left over from the previous pulse-driven convective episode), such as to produce $^{13}$C through $^{12}$C+$p$ (Goriely and Mowlavi (1998).

\begin{figure}[t]
\includegraphics[angle=-90,width=.45\textwidth]{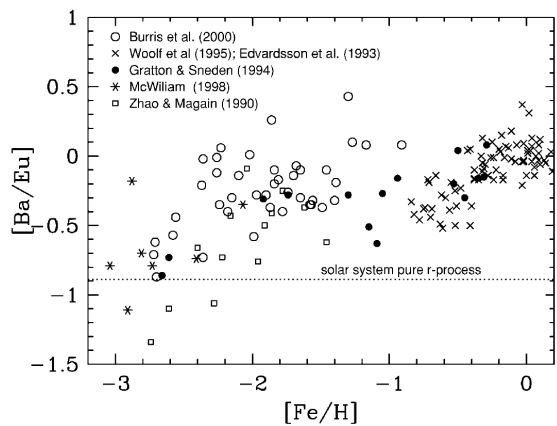}
\caption{Ba/Eu ratio as a function of metallicity, from various surveys (Truran et al. 2002). Note that the data of Burris et al. (2000) clearly show that the ratio rises above the expected solar system r-value at a metallicity [Fe/H]$\sim$-2.7. It is hard to understand such an early rise if the s-process Ba is produced in low mass AGB stars (see also next figure); massive AGBs might do it, but then the neutron source could not be $^{13}$C($\alpha,n$), contrary to the curent "paradigm".
}
\end{figure}

A generic result of such models is that all AGB stars, irrespectively of their metallicity, produce nuclei with 90$<$A$<$140 in an almost solar distribution, while only low metallicity stars ($Z<$0.01) may synthesize  heavier s-nuclei, like Pb. The reason for the latter property is that, as one goes to lower metallicities, the number of released neutrons {\it per} Fe seed nucleus increases (since the abundance of Fe decreases, while the abundance of $^{13}$C
remains the same, if production by $^{12}$C+$p$ is assumed). This property of the $^{13}$C($\alpha,n$) source  agrees well with recent observations of low metallicity stars rich in Pb (Van Eck et al. 2001). However, some modulation (by hand, at present) of the $^{13}$C abundance is required to explain other observations, concerning Pb poor stars of the halo (Aoki et al. 2002).

The mixing process of protons in the C-rich zones of low mass AGB stars is the principal unknown in studies of the production of s-nuclei in the framework of the current paradigm (see Goriely 2005 for a review of  recent ideas, like diffusive oveshoot, rotationally induced mixing or gravity waves). 
However, from the point of view of GCE, the situation is far from satisfactory.
Indeed, observations of neutron capture elements in low metallicity halo stars reveal an early rise of the Ba/Eu ratio (Burris et al. 2000 and Fig. 9). Eu is an r-process element, co-produced with Fe in core-collapse supernovae and appears quite early in the Galaxy. Ba has also a small r-component ($\sim$20\% in the solar system) but it is mostly produced by the main s-component of the s-process. If it originates in long-lived, low mass stars (where the $^{13}$C($\alpha,n$) source operates), then it is expected to arrive much later in the galactic scene, i.e. towards the end of the halo phase; this is found by quantitative GCE models exploring this standard senario (Travaglio et al. 1999),
as shown in Fig. 10. Taken at face value, the discrepancy between theory and observations implies that Low mass AGB stars do not produce the s-process Ba seen in old halo stars. A different site, involving more massive progenitors
(and $^{22}$Ne$\alpha,n$) as neutron source) should be seeked (see Goriely 2005 for such a possibility).

\begin{figure}[t]
\includegraphics[angle=-90,width=.45\textwidth]{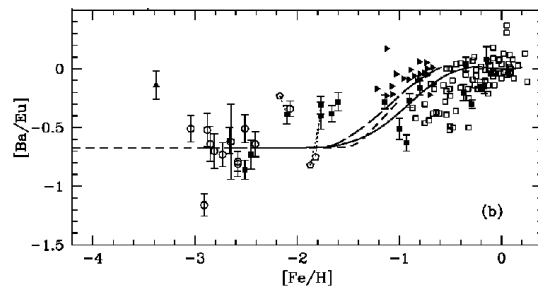}
\caption{Evolution of Ba/Eu ratio as a function of metallicity. Model predictions of Travaglio et al. (1999), based on production of s-nuclei mainly in low mass AGB stars, display a late rise of that ratio (around [Fe/H]$\sim$-1.2); they clearly do not reproduce the observations of the previous figure.
}
\end{figure}

%%%%%%%%%%%%%%%%%%%%%%%%%%%%%%%%%%%%%%%%%%%%%%%%
%% BACKMATTER
%%%%%%%%%%%%%%%%%%%%%%%%%%%%%%%%%%%%%%%%%%%%%%%%

\begin{theacknowledgments}
 I am indebted to the organisers, and to Grazyna Stasinska, in particular, for their invitation in that meeting and for financial support.
\end{theacknowledgments}



\bibliographystyle{aipproc}   % if natbib is available

\bibliography{sample}

\IfFileExists{\jobname.bbl}{}
 {\typeout{}
  \typeout{******************************************}
  \typeout{** Please run "bibtex \jobname" to optain}
  \typeout{** the bibliography and then re-run LaTeX}
  \typeout{** twice to fix the references!}
  \typeout{******************************************}
  \typeout{}
 }

%\end{document}

%%%%%%%%%%%%%%%%%%%%%%%%%%%%%%%%%%%%%%%%%%%
%% The following lines show an example how to produce a bibliography
%% without the help of the BibTeX program. This could be used instead
%% of the above.
%%%%%%%%%%%%%%%%%%%%%%%%%%%%%%%%%%%%%%%%%%%

\end{document}